\documentclass[fleqn,twoside]{article}
\usepackage{espcrc2}

\newcommand{\Einit}{E_{\mbox{\tiny init}}}

\usepackage{graphicx}
\usepackage[figuresright]{rotating}

\usepackage{amssymb} 

\newcommand{\AmS}{{\protect\the\textfont2
  A\kern-.1667em\lower.5ex\hbox{M}\kern-.125emS}}

\title{Generating Gravitational Waves After Inflation}

\author{Richard Easther\address{ Department of Physics, Yale University, New Haven CT 06520, United States of America. }%
        \thanks{RE is supported in part by the United States Department of Energy, grant DE-FG02-92ER-40704 and by an NSF Career Award PHY-0747868. I thank Richard Anantua, John T. Giblin and Eugene Lim for their collaboration on the research discussed here and this Proceedings borrows liberally from our papers \cite{Easther:2007vj,Anantua:2008am}.},
       }
       
\begin{document}
\begin{abstract}
 
 I review two mechanisms by which gravitational waves can be generated   at the {\em end\/} of inflation: preheating, and gravitons Hawking radiated during the decay of very small primordial black holes. These mechanisms are contrasted with the gravitational waves  {\em during\/} inflation, and may provide a window into the physical processes that govern the end of the inflationary phase.

 \vspace{1pc}
\end{abstract}

\maketitle

\section{INTRODUCTION}

It has long been known that an inflationary phase in the very early universe sources a background of gravitational waves. The amplitude of this background is tightly constrained by Cosmic Microwave Background [CMB] data \cite{Brown:2009uy,Komatsu:2008hk} placing nontrivial constraints on the inflationary parameter space \cite{Peiris:2008be}.    The temperature anisotropy of the CMB is measured with much higher accuracy than the polarization anisotropy (although constraints on the latter are rapidly improving \cite{Chiang:2009xs}) but ultimately polarization measurements will provide the strongest constraints on the primordial tensor background \cite{Baumann:2008aq}. The amplitude of this signal is strongly correlated with the physical scale of inflation, and there is no guarantee that inflation occurs at a high enough scale for this primordial signal to be observable, even in an ideal experiment.     In the longer term, space-based interferometers might make direct detection of this background at wavelengths of thousands or even millions of kilometers, but these comoving scales still left the horizon a number of e-folds before inflation ended.  

Recently, substantial attention has been paid to physical processes at the end of inflation which generate gravitational waves -- either preheating \cite{Easther:2007vj,Easther:2006gt,GarciaBellido:2007dg,Easther:2006vd,GarciaBellido:2007af,Dufaux:2007pt,Price:2008hq}, or Hawking radiation from primordial black holes \cite{Anantua:2008am}.   In contrast to the usual inflationary signal, these backgrounds are sharply peaked, and have frequencies corresponding to ``laboratory'' (or, at their lowest, solar system) scales, as opposed to the astrophysical scales associated with the CMB. Their amplitude can be substantial, but the technical challenges associated with the detection of high frequency gravitational waves suggest that it may be some time before these backgrounds are directly constrained.  Conversely, these backgrounds are sensitive to physical processes occurring at the end of inflation, and the coupling between the inflaton field(s) and the rest of the matter sector, and could provide a unique window into the primordial universe.

\section{GRAVITATIONAL WAVES FROM PREHEATING}
 
\begin{figure}[htb]
\includegraphics[width=7cm]{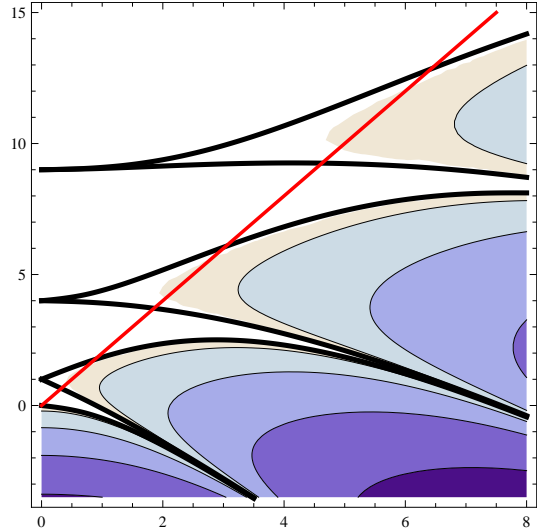}
\caption{ The imaginary part of the Mathieu critical exponent is plotted, with darker colors corresponding to a larger imaginary component.  Outside the heavy black lines the exponent is real-valued, and the corresponding solutions are strictly oscillatory.  The diagonal line corresponds to $A = 2 q$.\label{fig:stabilitychart}}
\end{figure}

Parametric resonance and preheating following inflation has been studied in great detail: see, for example \cite{Kofman:1994rk,Kofman:1997yn,Greene:1997fu,GarciaBellido:1997wm,Greene:1997ge,Bassett:2005xm} and references therein.  Preheating as a source of gravitational waves was first discussed by Khlebnikov and Tkachev \cite{Khlebnikov:1997di}, and the topic was revived by Easther and Lim \cite{Easther:2006gt} in 2006, who discussed the scaling relationships between the gravitational wave signal and the inflationary scale, confirmed in \cite{Easther:2006vd}. At least three other groups have developed codes for computing signal \cite{GarciaBellido:2007dg,GarciaBellido:2007af,Dufaux:2007pt,Price:2008hq}, which is a nontrivial numerical problem, and there is excellent agreement between them \cite{Price:2008hq}, although the internal workings of the codes differ significantly. 
Consider a toy model comprised of a {\it classical} inflaton $\phi$ coupled to a second field $\chi$ with Lagrangian
\begin{equation}
\mathcal{L} = \frac{1}{2}\partial_\mu \phi\partial^\mu \phi + \frac{1}{2}
\partial_\mu \chi \partial^\mu \chi -    \frac{1}{2}m^2 \phi^2  -    \frac{1}{2}g^2\phi^2\chi^2 \, .
\end{equation}
The equation of motion for $\chi$ is
\begin{equation}
\ddot{\chi} +3H\dot{\chi}-\frac{1}{a^2}\nabla^2\chi +   g^2\phi^2\chi = 0 \, .
\end{equation}
Expanding $\chi$ in terms of its momentum (Fourier) modes we obtain
\begin{equation}
\label{eomkmode}
\ddot{\chi}_k + 3H \dot{\chi}_k+\left(\frac{k^2}{a^2}+ g^2\phi^2\right)\chi_k = 0 \, .
\end{equation}
  As inflation ends, $\phi$ oscillates about zero.  
  If we ignore the expansion of the universe $H=0$  and $\phi$ oscillates with constant amplitude $\Phi$, although backreaction from $\chi$ particle creation  extracts energy from the $\phi$ field, eventually terminating resonance.  Changing variables to 
\begin{equation}
q = \frac{g^2\Phi^2}{4m^2}, \, A =   \frac{k^2}{m^2}+2q, \, 
z=mt,
\label{mathieuparameters}
\end{equation}
turns (\ref{eomkmode}) into a Mathieu Equation,
\begin{equation}
\chi_k^{\prime\prime} + (A-2q\cos(2z))\chi_k = 0.
\label{mathieueqn}
\end{equation}
where primes denote differentiation with respect to $z$.  Every solution to Mathieu's
equation has two parts,
\begin{equation}
\chi_k \propto f(z) e^{\pm i\mu z}
\end{equation}
where $f(z)$ is periodic and $\mu$ isthe {\it Mathieu characteristic exponent}. If $\mu$  has an
imaginary part the solution has an exponentially growing mode.  Figure \ref{fig:stabilitychart} shows the values of $\Im(\mu)$ as a function of $A$ and $q$.  We see that (\ref{mathieuparameters}) requires $A \ge 2q$, so we are interested in the parameter values that lie to the left of the diagonal line in Figure~\ref{fig:stabilitychart}. 

The amplification of selected $\chi_k$ (and, by backreaction, $\phi_k$)  makes the universe increasingly inhomogeneous, yielding an inhomogeneous, time dependent energy density -- which necessarily leads to the emission of gravitational radiation.    Ultimately,  the only way to follow the full evolution is via numerical simulation, and then extracting the contribution to the tensor modes.  

The stress-energy tensor associated with gravitational radiation is given by \cite{Misner:1974qy} 
\begin{equation}
T_{\mu\nu} = \frac{1}{32\pi G} \left<h_{ij,\mu}h^{ij}_{\,\,\,,\nu}\right>,
\end{equation}
and is specific to the {\it transverse-traceless} part of the metric
perturbation.  The $\langle \cdots \rangle$ denotes a spatial average and since we are solve for the field values (and $h_{ij}$) numerically, we simply integrate over the full ``grid'' on which our numerical solutions are computed.  The associated energy energy density is the $00$ component,
\begin{equation} 
\rho_{gw} = \frac{1}{32\pi G} \left<h_{ij,0}h^{ij}_{,0}\right> = \sum_{i,j}
\frac{1}{32 \pi G} \left<h^2_{ij,0}\right>. \label{gwdensity}
\end{equation} 
We compute $h_{ij}$ by extracting the source terms for the equations of motion obeyed by the  $g_{ij}$ from the full $T_{\mu\nu}$ obtained from numerical solutions of the scalar field dynamics \cite{Easther:2007vj}. We express the gravitational wave power in terms of its contribution to the fractional energy density in gravitational waves, $\Omega_{gw}$, per logarithmic interval in wavenumber,  
\begin{equation}
\label{omega}
\frac{d\Omega_{gw}}{d\ln k} = \frac{1}{\rho_{crit}}\frac{d \rho}{d \ln k} =
\frac{\pi k^3}{3H^2 L^2}\sum_{i,j}|h_{ij,0}(k)|^2.
\end{equation}
which we then convert in present-day units via an appropriate matching condition \cite{Easther:2007vj}.  

\begin{figure}
\begin{center}
\includegraphics[width=7cm]{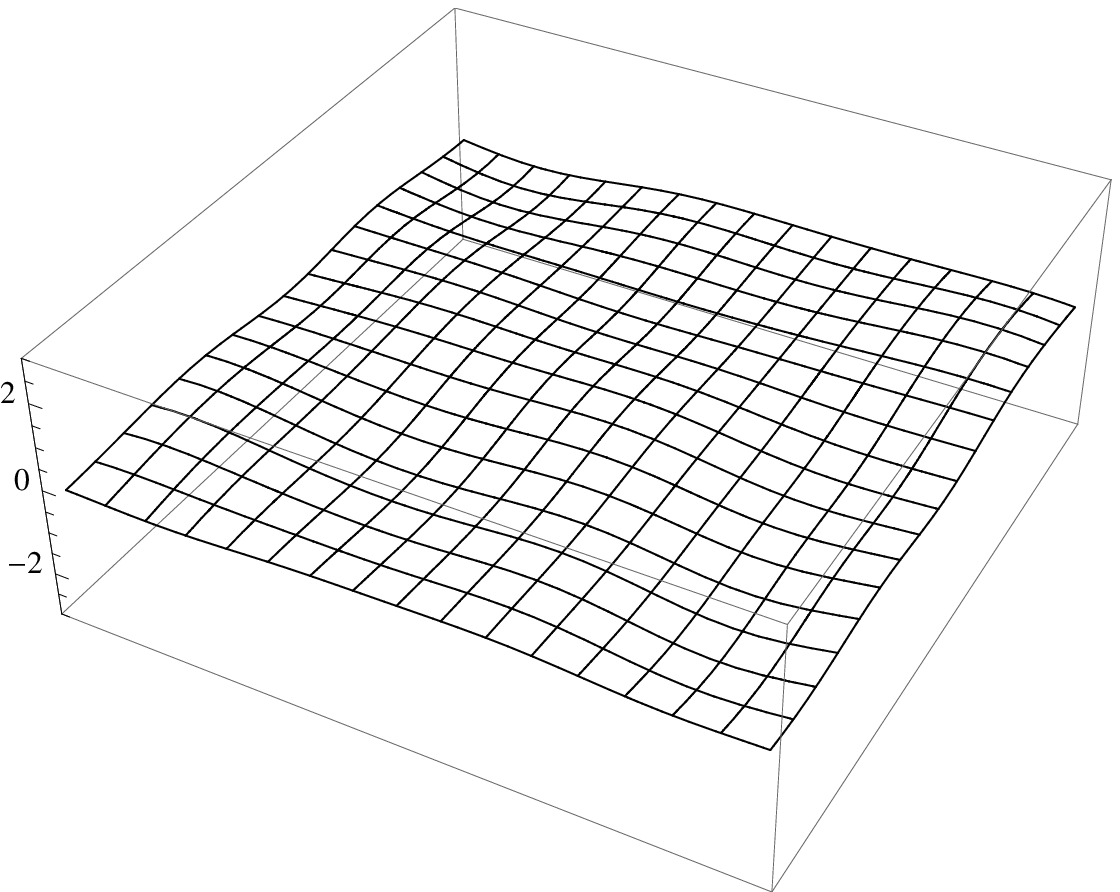}
\includegraphics[width=7cm]{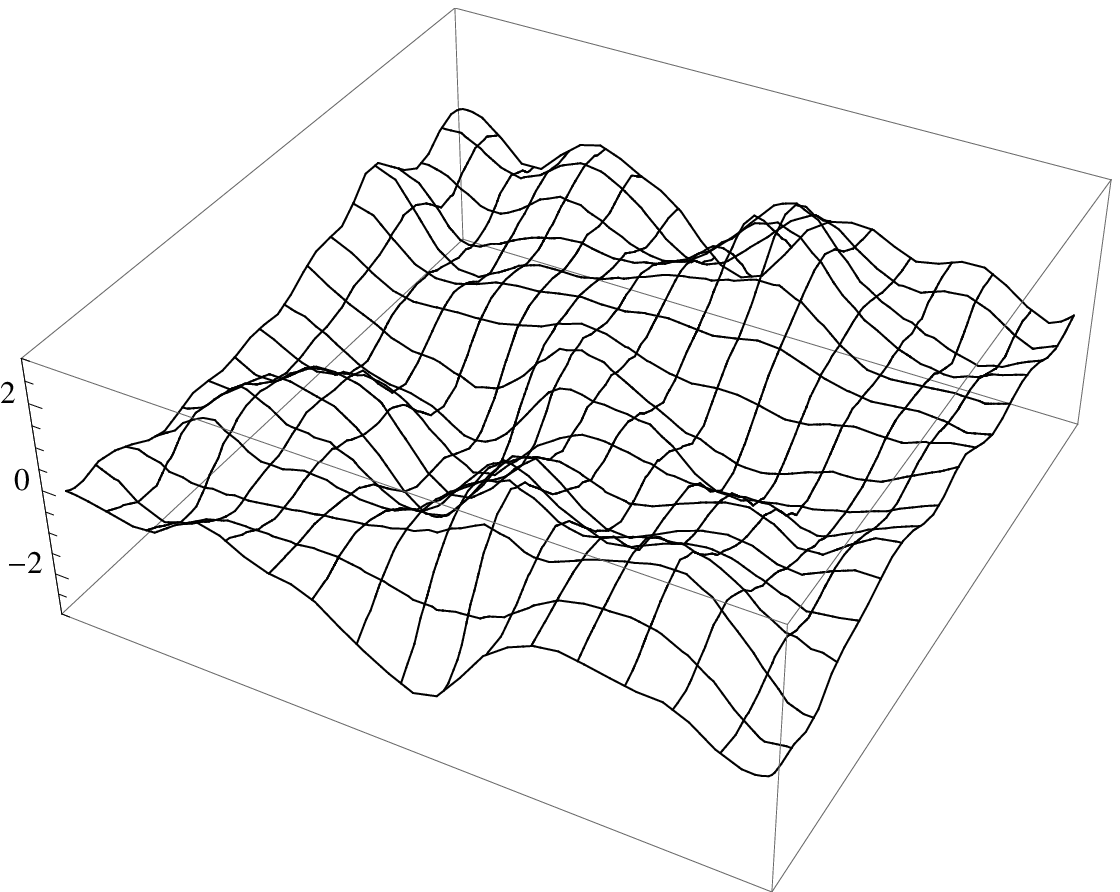}
\includegraphics[width=7cm]{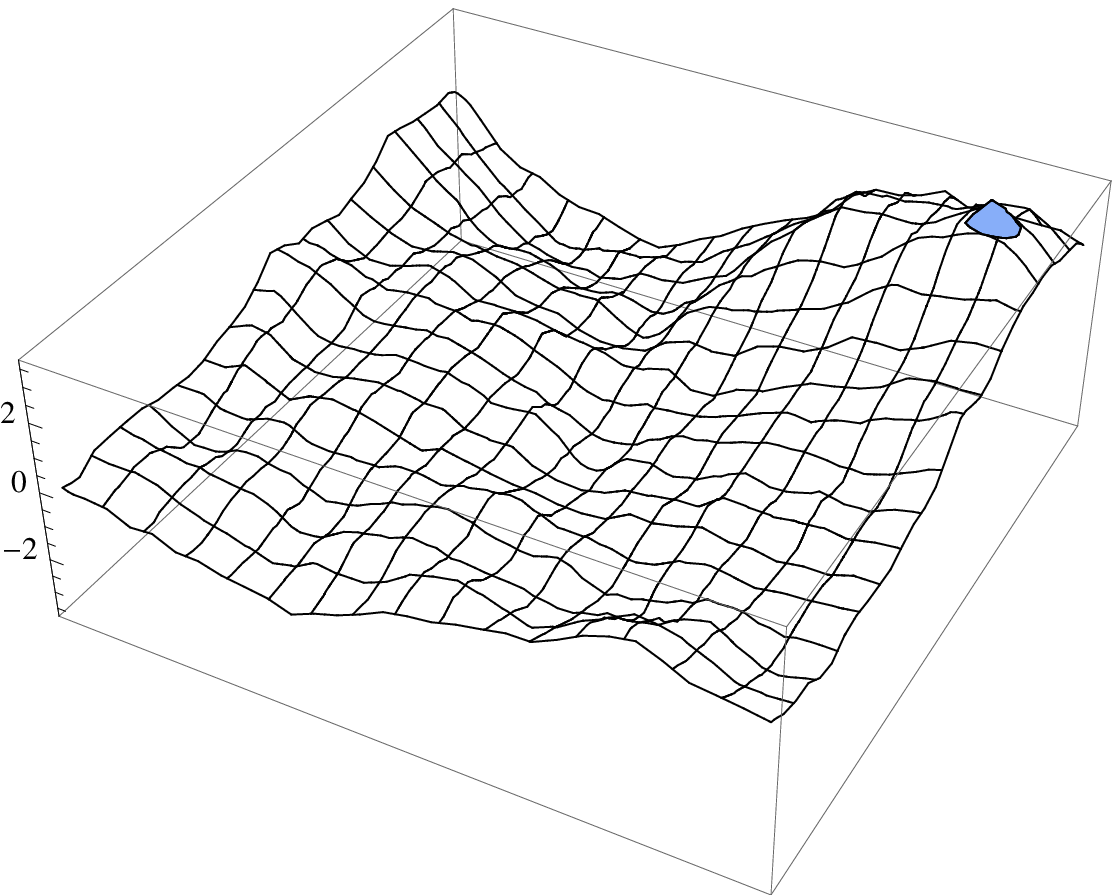}
\end{center}
\caption{A cross section of the $\chi$ field at three times, with an arbitrary normalization.  Gravitational waves are induced by (time varying) spatial gradients, which peak in the middle frame.\label{fig:field}}
\end{figure}

\begin{figure*}
\begin{center}
\includegraphics[width=10cm]{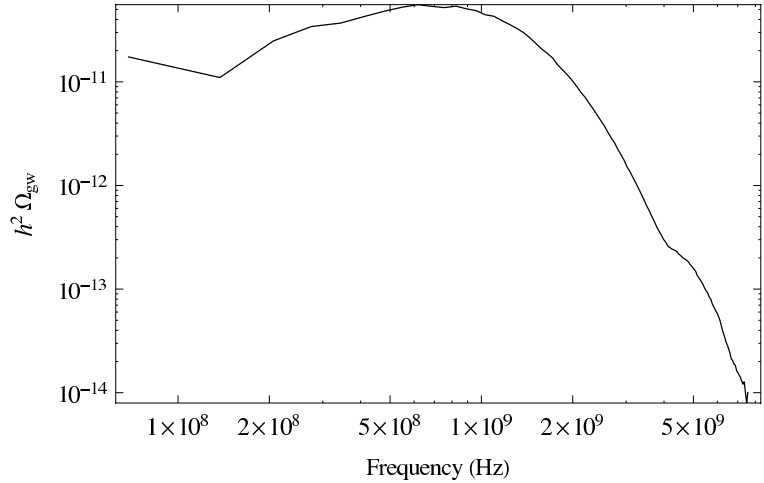}
 \end{center}
\caption{ The gravitational wave power (at the present epoch) generated with $m=10^{-6} m_p$ and $g^2 m_p^2 / m^2 =2.5 \times 10^5$. \label{fig:power}}
\end{figure*}

The resulting gravitational wave spectrum is sharply peaked, and located at very high frequencies when compared to both suspected astrophysical sources, and current and future interferometric experiments.      The gravitational wave background depends on the details of the resonance bands in the underlying cosmological model, and modes move in and out of resonance bands as the universe expands.  For the example considered here, the lowest lying resonance band is dominant. Modes whose physical wavenumber $k/a(t)$  never enter this region during preheating are not significantly amplified, resulting in the cutoff we see at high frequencies in Figure~\ref{fig:power}.   

In this model (and many resonant scenarios), the resonant modes are not vastly smaller than the post-inflationary Hubble scale. If the universe promptly reheats to the GUT scale  the comoving post-inflationary Hubble scale is mapped to centimeter scales in the present universe, and the spectrum peaks near GHz scales. If the inflationary scale is {\em lowered\/}, the signal appears at lower frequencies, because the growth of the universe between the present day and the preheating epoch is necessarily smaller \cite{Easther:2006gt} -- and if the energy scale is very low (closer to the TeV scale than the GUT scale), a background of gravitational waves generated during preheating would provide an alternative target for experiments like BBO.   

The amplitude of this signal does not necessarily change with the inflationary energy scale.  Only the gradient terms of the scalar field which source the gravitational wave background, and in many preheating scenarios the gradient energy rises until its average value is similar to that of the kinetic and potential energies. At fixed density this amplitude of the source terms are thus maximized.   Conversely, the {\em initial\/} power generated will decrease with the inflationary scale, since the quadruple and higher moments will be smaller. However,  the overall expansion of the universe between preheating and the present day is also reduced in these scenarios. The energy density in gravitational waves scales like radiation,  so $\rho){gw} \sim (a_{preheat} /a_0)^4$ -- and the combination of these two factors cancels, leading to a maximal amplitude of $\Omega_{gw} (k) \sim 10^{-10}$ -- a number also seen in bubble collision calculations \cite{Kosowsky:1991ua,Kosowsky:1992rz}.
  
 \section{PRIMORDIAL BLACK HOLES}
 
 A second, and more recently explored, mechanism which may generate gravitational waves in the post-inflationary universe is Hawking radiation from decaying Primordial Black Holes [PBH] \cite{Anantua:2008am}.   This analysis focusses on very small black holes, which decay before the onset of nucleosynthesis, and whose initial abundance is otherwise very weakly constrained.  It has long been  recognized that gravitons would be emitted as these black holes decay, although the precise rate is significantly modified by grey body corrections \cite{Page:1976df}.  
 
Assume a PBH  population whose mass is equal to the energy contained inside the Hubble volume at the instant they collapse.  Recalling that $H^2 = 8\pi \rho /3M_p^2$, and defining $\rho = \Einit^4$,  
\begin{equation}
M_{BH}  = \sqrt{\frac{3}{32 \pi}} \frac{M_p^3}{\Einit^2}
\end{equation}
which is the mass contained inside a sphere of radius $1/H$. Including grey body corrections $\Gamma_{sl}$  a Schwarzschild black hole emits (massless) particles with momentum $k$,  reducing its total energy as
\begin{eqnarray} \label{eqn:kt}
\frac{dE}{dtdk}&=& 
-\frac{2 g }{\pi } \frac{M_{BH}^2 }{M_p^4  }   \frac{k^3}{ e^{k/T} -1} \, \\
T &=& \frac{M_p^2}{8\pi M_{BH}} \, .
\end{eqnarray}
where    $g$ is the {\em effective\/} number of (bosonic) degrees of freedom, after integrating over grey body corrections \cite{Anantua:2008am}.  Here $k$ is the {\em physical\/} wavenumber  and the black hole  lifetime is
\begin{equation}  \label{eq:lifetime}
\tau   = \frac{10,240 \pi}{ g} \frac{M_{BH}^3}{M_p^4} = \frac{240}{g} \sqrt{\frac{3}{2\pi}} \frac{M_p^5}{\Einit^6} \, .
\end{equation}
An upper bound on $\Einit$ comes from the inflationary energy scale, which is constrained by the non-detection of a primordial gravitational wave background in the CMB. Since  we are interested in black holes which decay prior to nucleosynthesis with time to spare for thermalization, we need $\tau \lesssim 100$~s. With $\Einit = 10^{12}$~GeV,   $\tau \approx 29/g$~s, and the initial temperature is  $18.8$~TeV.  Standard model states alone give $g\sim {\cal O}(10^2)$.

\begin{figure}[tb]
\includegraphics[width=3in]{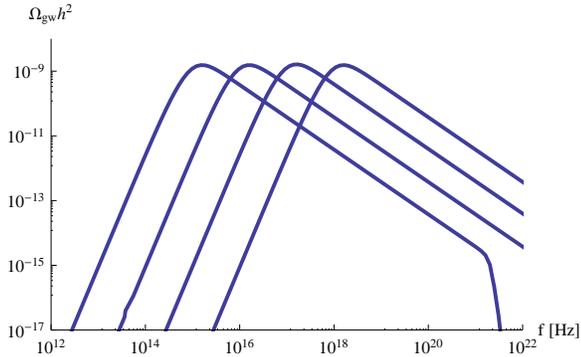}  
\caption{\label{fig:E}   $\Omega_{gw}(f)h^2$ with (from left to right) $\Einit =  10^{15}, 10^{14}, 10^{13}$, and $10^{12}$ GeV. In all cases $\beta=0.001$ and $g=1000$. }
 \end{figure}
   The mass-fraction of PBH is denoted $\Omega_{BH}$. Initially $\Omega_{BH}   = \beta$. We assume $\beta \ll 1$ and that the remaining matter consists of radiation.    Denoting the number density of PBH by $n(t)$, the energy density $\rho_{BH} = n(t) M_{BH}(t)$ we get the following set of equations
\begin{eqnarray}
\frac{d \rho_{BH}}{dt} &=& \dot{n}(t) M_{BH} + n(t) \dot{M}_{BH} \, , \\
&=& -3 \frac{\dot{a}}{a} \rho_{BH} + \rho_{BH} \frac{\dot{M}_{BH}}{M_{BH}} \, ,\\
\frac{ d \rho_{rad}}{dt} &=& -4\frac{\dot{a}}{a}\rho_{rad} - \rho_{BH} \frac{\dot{M}_{BH}}{M_{BH}}  \, ,\\
\frac{\dot{a}}{a} &=& \left[\frac{8 \pi }{3 M_p^2} (\rho_{BH}+ \rho_{rad}) \right]^{1/2} \,  
\end{eqnarray}
along with equation~(\ref{eqn:kt}) from which one obtains $\Omega_{gw}(k)$ by an appropriate rescaling. 

We again express our results in terms of the present-day spectral energy density of gravitational radiation \cite{Easther:2006gt,Price:2008hq}.   Figure~\ref{fig:E}  shows the present-day power in gravitational radiation as a function of $\Einit$. The gravitational wave power is substantial, and found at very high frequencies, relative both to any preheating signal, and the electromagnetic radiation in the CMB.   Roughly speaking, the temperature of the universe scales as $1/a(t)$. A decaying black hole is much hotter than the surrounding universe, but the emitted gravitational waves are redshifted by the same factor as other radiation. Consequently, these gravitational waves will necessarily have a higher frequency than the present-day CMB, which peaks around $100$~GHz. Lowering $\Einit$ increases the PBH lifetime, enhancing this discrepancy and pushing the gravitational wave signal to higher (present day) frequencies.    The signal  also depends on $g$ and  $\beta$, as explained in \cite{Anantua:2008am}.  
 
These primordial black holes survive for many Hubble times, and since their density scales like matter, even a small initial population can lead to a universe that undergoes a transient period of matter domination before the black holes decay, which then rethermalizes the universe. The additional phenomenology arising from any such matter dominated stage has not been fully explored, and remains a topic for future enquiry.

 \section{DISCUSSION}

I  have reviewed two different mechanisms for generating gravitational waves at the end of inflation -- via parametric resonance or preheating, and graviton emission from decaying primordial black holes.  Physically these mechanisms are very different -- the first is  driven by the higher moments of the matter distribution and thus operates in the purely classical limit. Conversely, Hawking radiation is intrinsically quantum mechanical, and would not take place if  $\hbar \equiv 0$.  However, both mechanisms depend critically on the detailed behavior of the universe makes the transition between inflationary and ``regular'' expansion. Parametric resonance and preheating depends sensitively on the couplings between the inflaton and other particles, whereas primordial black holes only form in models with a sharp rise in the primordial perturbation spectrum as inflation comes to an end.

In most cases, the frequencies of these gravitational wave backgrounds far exceed those which modern experiments such as LIGO or VIRGO, and space-based successors such as LISA, are designed to detect.  Consequently, if meaningful constraints are to be placed on the backgrounds, new detector technologies are likely to be required. The one exception to this statement is if preheating occurs after low scale (ie TeV) inflation, in which case the gravitational wave signal could be visible in proposed space-based interferometric detectors. This is particularly relevant to BBO, which has been designed to detect the primordial gravitational wave signal generated during GUT scale inflation. This signal is invisible in low scale models, and the preheating signal provides an alternative source for a stochastic background, widening the set of inflationary models a BBO style experiment would constrain.

Finally, it is clear that these signals are all ``futuristic'' in that  placing meaningful constraints on their amplitude would require second or third generation gravitational wave experiments, and at the time of writing no direct gravitational wave detections have been reported.  However, from a philosophical standpoint they serve to underline the rich phenomenology that can be associated with inflationary models, and the importance of fully exploring their consequences and possible observable fingerprints. 
 


\begin{thebibliography}{10}

\bibitem{Easther:2007vj}
R.~Easther, J.~T. Giblin, and E.~A. Lim,
\newblock Phys. Rev. {\bf D77}, 103519 (2008), 0712.2991.

\bibitem{Anantua:2008am}
R.~Anantua, R.~Easther, and J.~Giblin, John~T.,
\newblock (2008), 0812.0825.

\bibitem{Brown:2009uy}
QUaD, M.~L. Brown {\em et~al.},
\newblock (2009), 0906.1003.

\bibitem{Komatsu:2008hk}
WMAP, E.~Komatsu {\em et~al.},
\newblock Astrophys. J. Suppl. {\bf 180}, 330 (2009), 0803.0547.

\bibitem{Peiris:2008be}
H.~V. Peiris and R.~Easther,
\newblock JCAP {\bf 0807}, 024 (2008), 0805.2154.

\bibitem{Chiang:2009xs}
H.~C. Chiang {\em et~al.},
\newblock (2009), 0906.1181.

\bibitem{Baumann:2008aq}
CMBPol Study Team, D.~Baumann {\em et~al.},
\newblock (2008), 0811.3919.

\bibitem{Easther:2006gt}
R.~Easther and E.~A. Lim,
\newblock JCAP {\bf 0604}, 010 (2006), astro-ph/0601617.

\bibitem{GarciaBellido:2007dg}
J.~Garcia-Bellido and D.~G. Figueroa,
\newblock Phys. Rev. Lett. {\bf 98}, 061302 (2007), astro-ph/0701014.

\bibitem{Easther:2006vd}
R.~Easther, J.~Giblin, John~T., and E.~A. Lim,
\newblock Phys. Rev. Lett. {\bf 99}, 221301 (2007), astro-ph/0612294.

\bibitem{GarciaBellido:2007af}
J.~Garcia-Bellido, D.~G. Figueroa, and A.~Sastre,
\newblock Phys. Rev. {\bf D77}, 043517 (2008), 0707.0839.

\bibitem{Dufaux:2007pt}
J.~F. Dufaux, A.~Bergman, G.~N. Felder, L.~Kofman, and J.-P. Uzan,
\newblock Phys. Rev. {\bf D76}, 123517 (2007), 0707.0875.

\bibitem{Price:2008hq}
L.~R. Price and X.~Siemens,
\newblock Phys. Rev. {\bf D78}, 063541 (2008), 0805.3570.

\bibitem{Kofman:1994rk}
L.~Kofman, A.~D. Linde, and A.~A. Starobinsky,
\newblock Phys. Rev. Lett. {\bf 73}, 3195 (1994), hep-th/9405187.

\bibitem{Kofman:1997yn}
L.~Kofman, A.~D. Linde, and A.~A. Starobinsky,
\newblock Phys. Rev. {\bf D56}, 3258 (1997), hep-ph/9704452.

\bibitem{Greene:1997fu}
P.~B. Greene, L.~Kofman, A.~D. Linde, and A.~A. Starobinsky,
\newblock Phys. Rev. {\bf D56}, 6175 (1997), hep-ph/9705347.

\bibitem{GarciaBellido:1997wm}
J.~Garcia-Bellido and A.~D. Linde,
\newblock Phys. Rev. {\bf D57}, 6075 (1998), hep-ph/9711360.

\bibitem{Greene:1997ge}
B.~R. Greene, T.~Prokopec, and T.~G. Roos,
\newblock Phys. Rev. {\bf D56}, 6484 (1997), hep-ph/9705357.

\bibitem{Bassett:2005xm}
B.~A. Bassett, S.~Tsujikawa, and D.~Wands,
\newblock Rev. Mod. Phys. {\bf 78}, 537 (2006), astro-ph/0507632.

\bibitem{Khlebnikov:1997di}
S.~Y. Khlebnikov and I.~I. Tkachev,
\newblock Phys. Rev. {\bf D56}, 653 (1997), hep-ph/9701423.

\bibitem{Misner:1974qy}
C.~W. Misner, K.~S. Thorne, and J.~A. Wheeler,
\newblock San Francisco 1973, 1279p.

\bibitem{Kosowsky:1991ua}
A.~Kosowsky, M.~S. Turner, and R.~Watkins,
\newblock Phys. Rev. {\bf D45}, 4514 (1992).

\bibitem{Kosowsky:1992rz}
A.~Kosowsky, M.~S. Turner, and R.~Watkins,
\newblock Phys. Rev. Lett. {\bf 69}, 2026 (1992).

\bibitem{Page:1976df}
D.~N. Page,
\newblock Phys. Rev. {\bf D13}, 198 (1976).

\end{thebibliography}
\end{document}